\documentclass[aps,prl,twocolumn]{revtex4}
\usepackage{amsfonts}
\usepackage{amsmath}
\usepackage{amssymb}
\usepackage{graphicx}

\input{tcilatex}
\begin{document}

\title{Calculated Momentum Dependence of Zhang-Rice States in Transition
Metal Oxides}
\author{Quan Yin$^{1}$, Alexey Gordienko$^{1,2}$, Xiangang Wan$^{1,3}$,
Sergey Y. Savrasov$^{1}$}
\affiliation{$^{1}$Department of Physics, University of California, Davis, CA 95616}
\affiliation{$^{2}$Department of Physics, Kemerovo State University, Kemerovo, Russia}
\affiliation{$^{3}$National Laboratory of Solid State Microstructures and Department of
Physics, Nanjing University, Nanjing, China}

\begin{abstract}
Using a combination of local density functional theory and cluster exact
diagonalization based dynamical mean field theory, we calculate many body
electronic structures of several Mott insulating oxides including undoped
high T$_{c}$ materials. The dispersions of the lowest occupied electronic
states are associated with the Zhang-Rice singlets in cuprates and with
doublets, triplets, quadruplets and quintets in more general cases. Our
results agree with angle resolved photoemission experiments including the
decrease of the spectral weight of the Zhang--Rice band as it approaches $%
\mathbf{k}$=0.
\end{abstract}

\date{\today }
\maketitle

Quasiparticle excitations in insulating transition metal oxides (TMOs)\ such
as classical Mott--Hubbard systems or undoped high temperature
superconductors (HTSCs) have been puzzling electronic structure theorists
for many years \cite{Fujimori,Pickett}. While photoemission experiments in
these materials show \cite{ARPESreview} the existence of the $d$--states
located both right below the Fermi energy and at much higher binding
energies (typically $\sim $10 eV), it is difficult to understand this
genuine many--body redistribution of the spectral weight using calculations
based on a static mean field theory \cite{Oguchi,Jepsen}, such as the
density functional theory (DFT) in its local density approximation (LDA) 
\cite{DFTreview}. Modern approaches, such as LDA+U \cite{LDA+U}, can
differentiate between charge--transfer and Mott--Hubbard natures of these
systems \cite{ZSA}, but still have difficulties in recovering insulating
behavior of the paramagnetic (PM) state and tackling more complicated
many--body features such as Zhang-Rice (ZR) singlet of HTSCs \cite%
{ARPESreview,ZRS}. Only most recent developments based on a combination of
local density approximation (LDA) and dynamical mean field theory (DMFT) 
\cite{RMP2006} have started to address those issues \cite{LDA++,Jeys,Kunes}.

In the present work, using a novel implementation of LDA plus cluster exact
diagonalization based DMFT we demonstrate how to obtain accurate spectra of
transition metal oxides and, in particular, describe full momentum dependent
low--energy excitations associated in those systems with antiferromagnetic
(AFM)\ Kondo--like coupling between a spin of oxygen hole injected by
photoemission process and a local magnetic moment of the transition metal
ion. These narrow energy bands are composed from the well known Zhang--Rice
singlet states in cuprates \cite{ZRS}, which have recently renewed their
attention in connection with the disappearance of their spectral weight as
the wave vector approaches the Brillouin Zone (BZ) center, and the observed
high energy kink entitled as "waterfall" feature \cite{Waterfalls}.
Zhang--Rice doublets have been discussed in NiO \cite{Zaanen}, and their
further generalizations to triplets (CoO), quadruplets (FeO) and quintets
(MnO) all naturally emerge from our LDA+DMFT\ calculations. We find that the
ZR states exhibit a similar behavior in all systems including the loss of
their spectral weight at the $\Gamma$ point, which can be understood as the
lack of hybridization between transition metal $d$ states and neighboring
oxygen $p$ states, the effect most pronounced in HTSCs. There is a generally
good agreement between our results and angle resolved photoemission (ARPES)
experiments.

In our self--consistent LDA+DMFT calculations, the LDA one--electron
Hamiltonians are supplemented with the self--energies of the $d$--electrons
that are extracted from the cluster exact diagonalizations involving a $d$%
--shell of a transition metal ion hybridizing with nearest oxygen orbitals.
We also include the effect of inter--site self--energies in the study of
cuprates including Sr$_{2}$CuO$_{2}$Cl$_{2}$ and La$_{2}$CuO$_{4}$ by
considering clusters of two copper $3d$ orbitals. The hybridization
functions in our calculations are fit using a single pole approximation $%
\Delta (\omega)=V^{2}/(\omega-E_{p})$ which delivers the effective position
of the bath level $E_{p}$ (interpreted as the oxygen $p$ band), as well as
its hybridization $V$ with the $d$ states. The cluster self--energies are
fit into a rational form $\Sigma(\omega)=\Sigma(\infty)+\sum_{i}W_{i}/(%
\omega-P_{i})$ using interpolation by three poles \cite{Am}. Such fits allow
to simplify the process of extracting parameters for the cluster
Hamiltonians and performing full self--consistency using the LDA+DMFT for
the actual materials considered in our work. Conventional values for the
Coulomb interaction matrices, which were found from constrained density
functional calculations \cite{LDA+U}, are used to perform the calculations
at temperatures above the long--range magnetic order.

Since dimensions of the many--body Hamiltonians quickly become prohibitively
large to handle by standard diagonalization algorithms, \ we have newly
implemented a kernel polynomial method (KPM)\ for extracting spectral
functions in our cluster calculations \cite{KPM}. Similar to the Lanczos
method, the KPM is an iterative procedure which allows to recover moments of
many body densities of energy states. Using 200 moments or so, the
one--electron Green's functions are found to converge accurately for our
cluster Hamiltonians with the dimensions up to 100,000.

We first discuss a long--standing problem of low energy excitations in
HTSCs. Early LDA calculations were unable to reproduce even the magnetic
behavior of those systems \cite{Pickett}. Although this problem was later
solved by the LDA+U\ method \cite{LDA+U}, these calculations miss the
important ZR\ physics and are unable to address the low--energy scale. On
the other hand, $t-J$ and $t-t^{\prime }-t^{\prime \prime }-J$ model
calculations have successfully reproduced the ZR dispersion as compared to
the ARPES data \cite{t-J of 2122,2122 data-circle,2122 data-triangle,2122
data-square,ARPES HTc theory}. A hot discussion has appeared in the
literature around most recent ARPES experiments which have shown that the ZR
band vanishes as the wave vector approaches the $\Gamma $ point, with the
spectral weight seemingly transferring to higher--energies as the
"waterfall" \cite{Waterfalls}. While it could be simple photoemission matrix
element effect, most recent dynamical cluster approximation (DCA) based
simulations of the 2D Hubbard model show \cite{Alexandru} large imaginary
part of the self--energy, resulting in asymmetric spectral functions
resembling the waterfall. As we generally see, the HTSCs still represent a
challenge for the theory, particularly for modern electronic structure
calculations, which try to incorporate all hopping integrals accurately and
use realistic values of the Coulomb interaction parameters.

\begin{figure}
[ptb]
\begin{center}
\includegraphics[height=2.8543in,width=3.4786in]%
{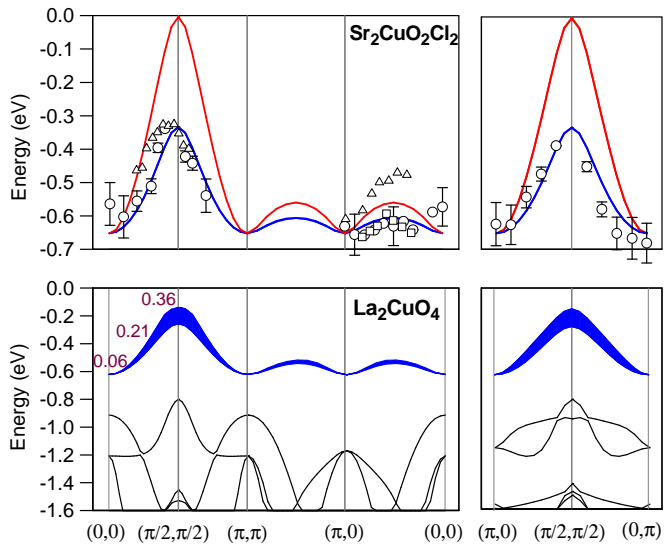}%
\caption{(color online) Calculated low energy
excitations in Sr$_{2}$CuO$_{2}$Cl$_{2}$ (top) and La$_{2}$CuO$_{4}$
(bottom) using LDA plus single--site DMFT (red line) and two--site cluster
DMFT (blue line). The experimental data are from: Ref. \protect\cite{2122
data-circle} (open circles), Ref.\protect\cite{2122 data-triangle} (open
triangles), Ref.\protect\cite{2122 data-square} (open squares). The blue
linewidth shows the oxygen content, while the numbers indicate the amount of
electrons in the ZR band.}%
\label{HTS}%
\end{center}
\end{figure}

In the present work LDA+DMFT calculations are performed for Sr$_{2}$CuO$_{2}$%
Cl$_{2}$, whose ARPES experiments clearly identified the ZR band \cite%
{ARPESreview}, and for La$_{2}$CuO$_{4}$. First, the calculations were done
for a single impurity case when the cluster is made of a copper ion
surrounded by 4 oxygen orbitals. This result is presented in Fig.\ref{HTS}%
(top panel) by red line together with various photoemission experiments
denoted by symbols. The dispersions of the ZR band in Sr$_{2}$CuO$_{2}$Cl$%
_{2}$ correctly follows the ARPES data, although its bandwidth is
overestimated by a factor of two, especially along $(0,0)$-$(\pi ,\pi )$
direction. The situation is somewhat unclear along $(0,0)$-$(\pi ,0)$ line
since experiments by two groups show quite different bandwidths (triangles%
\cite{2122 data-triangle} vs circles\cite{2122 data-circle} and squares\cite%
{2122 data-square}). From the standpoint of the single--impurity model, it
is however clear that the dispersion of the ZR band above the N\'{e}el
temperature is primarily governed by the short--range antiferromagnetic
correlations in that once the hole is moving in the lattice it is dressed up
if the short--range order is present. This effect is missing when the
single--impurity approximation is adopted as the DMFT treats the PM regime
as completely disordered state, resulting in the ZR band being too wide in
our calculation. To include short range magnetic correlations, we
subsequently performed calculations with two impurities by considering a
cluster of two copper atoms surrounded by 7 oxygens (one shared orbital).
The cluster exact diagonalization now delivers both the on--site $\Sigma
_{11}(\omega )$ and inter--site $\Sigma _{12}(\omega )$ self--energies,
which can be Fourier transformed to the form $\Sigma (\mathbf{k,}\omega
)=\Sigma _{11}(\omega )+\Sigma _{12}(\omega )(\cos k_{x}+\cos k_{y})$ and
acquire the much needed k--dependence. By fixing the frequency $\omega $ to
the position of the ZR band we are now able to add $\Sigma (\mathbf{k,}%
\omega )$ to the LDA Hamiltonian. The result of such a calculation is
depicted by the blue lines in Fig.\ref{HTS}. A remarkable band narrowing now
occurs where the ZR bandwidths come about 0.3 eV for Sr$_{2}$CuO$_{2}$Cl$%
_{2} $ and 0.4 eV for La$_{2}$CuO$_{4}.$ Using the calculated ZR dispersion,
the parameters for $t-J$ model are also evaluated. For Sr$_{2}$CuO$_{2}$Cl$%
_{2}$ they appear to be $J=0.135$eV, $t=0.34$eV, $t^{\prime }=-0.12$eV and $%
t^{\prime \prime }=0.08$eV, and for La$_{2}$CuO$_{4}$ the numbers are $%
J=0.175$eV, $t=0.44$eV, $t^{\prime }=-0.15$eV, $t^{\prime \prime }=0.10$eV.
The reduction of the ZR bandwidths by including k--dependence of the
self--energy demonstrates the importance of going beyond the single--site
approximation in describing the low--energy excitations in cuprates.

We further discuss the effect of disappearance of the ZR\ spectral weight
around the $\Gamma$ point. In Fig.\ref{HTS}, the blue fat line on the bottom
panel shows the amount of oxygen p states presented in the ZR band of La$%
_{2} $CuO$_{4}$. It is clearly seen that the spectral weight gets smoothly
reduced as the \textbf{k} vector approaches the BZ center. The same effect
is found in our calculations for other HTSCs. The reduction can be simply
understood using the 3--band tight--binding argument: the k--dependent
hybridization between $d_{x^{2}-y^{2\text{ }}}$and $p_{x},p_{y}$ orbitals is
exactly equal to zero at $\Gamma$, forbidding the formation of the ZR
singlet. We can also track this effect with $\Sigma(\mathbf{k},\omega)$
using a recently proposed matrix expansion algorithm \cite{Am}: Assuming
that two poles of the self--energy are needed to reproduce both the Hubbard
and the ZR bands, the resulting Hamiltonian is $5\times5$ in the extended
pole space (3 for $d_{x^{2}-y^{2\text{ }}},p_{x},p_{y}$\ and 2 for the
self--energy poles). It can be simply diagonalized and the same effect is
monitored. The amount of electrons in each band is less than one since
auxiliary pole states are added to the Hamiltonian \cite{Am}. Fig.\ref{HTS}
shows the actual number of electrons in the ZR band as a function of k,
decreasing towards $\Gamma$. Here, all spectral weight gets transferred to
the Hubbard bands located at the scale of $U$ and also to the oxygen band.

While we are unable to account for the effect of the imaginary self--energy
in the exact diagonalization for small clusters, recent DCA simulations of
the Hubbard model \cite{Alexandru} have shown very asymmetric k--dependent
spectral functions resembling the waterfall. On top of that a lot of other
states are present below $-1$ eV, as shown in Fig.\ref{HTS} (bottom panel).
In our opinion this provides most natural interpretation of the recent ARPES
data. As small hole doping occurs, the Fermi level sits at top of the ZR
band which would again contradicts with the standard LDA like description of
the cuprates, although we are unable to study the regime of optimal doping
by the present procedure requiring too large clusters to diagonalize.

\begin{figure}
[ptb]
\begin{center}
\includegraphics[height=2.9359in,width=3.477in]%
{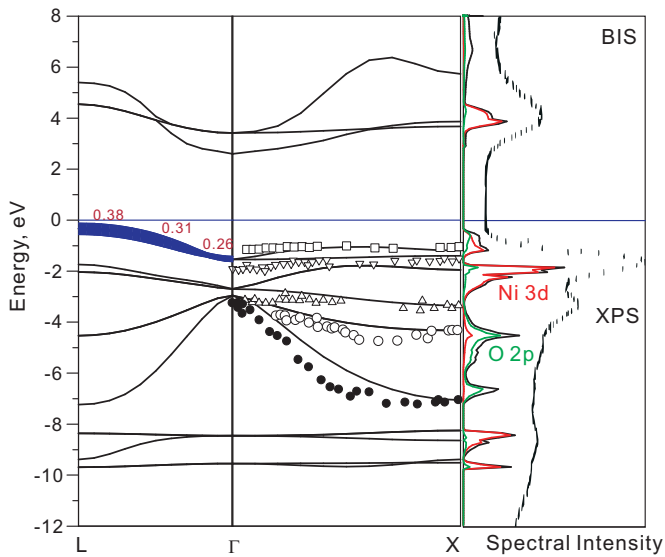}%
\caption{(color online) Comparison
between calculated quasiparticle dispersions using LDA+DMFT and the
photoemission data \protect\cite{Jepsen,NiO expt1} for paramagnetic state of
NiO. The blue linewidth and the numbers show the oxygen content and the
amount of electrons in the ZR band. }%
\label{NiO}%
\end{center}
\end{figure}

Let us now turn to the discussion of NiO which has been widely chosen as
another classical example of a strongly correlated system. Two decades ago
Fujimori \textit{et al} \cite{NiO CI} and Sawatzky \textit{et al} \cite{NiO
expt1} have shown that the valence band photoemission spectrum of NiO can be
understood from a configurational interaction (CI) approach, indicating that
the electronic properties here are local. The $3d$ electrons repulsion gives
rise to the lower and upper Hubbard bands located at $-9$ and $4$ eV,
respectively. The $3d-2p$ hybridization is responsible for the $3d$%
--character peak just below the Fermi level \cite{Mixer}. The discussion of
a generalized spin--fermion model derived for a slab of NiO shows a
well--developed set of ZR states which agree reasonably well within the
low--energy features seen in the ARPES experiments \cite{Zaanen}. However,
NiO exhibits both correlation effects and band--structure effects, plus
antiferromagnetic order under the N\'{e}el temperature of 520\ K. Below T$%
_{N}$ extensive studies of NiO have been performed using the LDA+U method,
and the positions of the Hubbard bands have been predicted correctly when
using the values of on-site Hubbard interaction $U=8$ eV \cite{LDA+U}.
Unfortunately, much of the $d$--electron spectral weight just below the
Fermi level is found to be lacking in this type of calculation unless the
value of $U$ is reduced to $4\sim5$ eV making this system of Mott--Hubbard
type rather than of charge--transfer type. Better than LDA+U, by using
atomic (Hubbard I) approximation for the self-energy, one can get insulating
solution even for paramagnetic case\cite{LDA++}. Also, recent comparisons
between LDA\ calculations of NiO in its paramagnetic (predicted to be
metallic) state and the ARPES experiments have indeed revealed \cite{Jepsen}
a rather good agreement between each other in the directions of BZ where the
LDA\ band dispersions remain insulating. One thus faces a dilemma of
matching different portions of the photoemission data with three different
types of electronic structure calculations (LDA+U with two values of U as
well as straight LDA).

We now show that all these problems can be overcome by performing LDA+DMFT
calculations. Fig.\ref{NiO} shows the comparison between calculated
quasiparticle dispersions and photoemission data for NiO \cite{Jepsen,NiO
expt1}. The Ni $3d$ spectral weight is redistributed between the lower
Hubbard band located at $-8\sim -10$ eV, upper Hubbard band located at
around $4$ eV, and strong peaks located just below the Fermi level. Having
the ground state in NiO as $|p^{6}d^{8}\rangle ,$ the lowest excitation here
is the result of the interaction between the spin of the oxygen hole and the
two hole $S=1$ state in the $d$--shell. The total spin of such cluster can
be $1/2$ (doublet) or $3/2$ (quadruplet). The antiparallel Kondo--like
coupling implies that the doublet state is lower in energy, in accord with
early predictions using the spin--fermion model \cite{Zaanen}. The situation
is akin to the singlet and triplet situation in cuprates. The direct energy
gap is seen in Fig.\ref{NiO} to be around 3.5 eV in our calculation (the
experimental value is 4 eV), although most of the optical transitions would
start at the energies 4 eV or so as is evident from our density of states
plots. We also see that both the positions and the dispersion of the oxygen
bands located at the energies $-4\sim -7$ eV below the Fermi level agree
well with the ARPES data.

It is remarkable that similar to the HTSCs the reduction of the spectral
weight of the ZR doublet is also seen in NiO when the k vector approaches
the $\Gamma$ point. Here the same tight--binding argument applies: in the
nearest neighbor approximation the k--dependent hybridization between oxygen 
$p$ and $d_{e_{g}}$ orbitals vanishes prohibiting the creation of the ZR
state. The amount of actual $p$ electrons in the ZR state is shown by
fattening this band in Fig. \ref{NiO}. While it decreases when $\mathbf{k}%
\rightarrow0$, there is some residual amount around $\Gamma$ due to other
hoppings which results in the actual number of electrons (shown by numbers)
to be not as drastically reduced at $\Gamma$ as in the case of HTSCs.

\begin{figure}
[ptb]
\begin{center}
\includegraphics[height=2.8383in,width=3.477in]%
{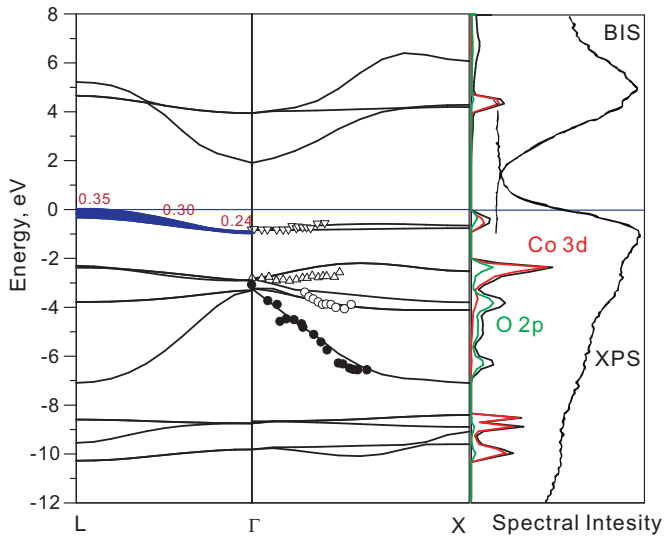}%
\caption{(color online) The same as Fig. 
\protect\ref{NiO} but for CoO. Photoemission data are from Ref. \protect\cite
{CoO ARPES,CoO XPS-BIS}}%
\label{CoO}%
\end{center}
\end{figure}

Our result for CoO demonstrates very similar behavior to that of NiO. Fig.%
\ref{CoO} compares the calculated quasiparticle dispersions with the
photoemission data \cite{CoO ARPES,CoO XPS-BIS}. The calculated energy gap
is $\sim $2.5 eV, and the experimental value is about 2.4 eV. Remarkably
similar to NiO, the $d$--electron spectral weight is distributed between the
Hubbard bands found at energies $-9\sim -10$ eV and at $4$ eV, as well as
the ZR band just below the Fermi level. The lowest excitation here can be
described as a spin triplet state because of the antiferromagnetic
interaction between the oxygen hole and the 3 hole $S=3/2$ state of the Co
ion. We also see that the spectral weight in the ZR\ triplet band gets
reduced as one moves towards the BZ center exactly as in the previous cases.%

We finally mention our calculations for FeO and MnO where a very similar
behavior was found. Since the spin momenta on Fe and Mn sites are roughly
equal to $2$ and $5/2$, respectively, the corresponding generalizations of
the ZR states are spin $3/2$ (quadruplet) for FeO, and spin $2$ (quintet)
for MnO. We should however mention that moving towards middle of the $3d$%
--metal oxide series, the hybridization effects become more and more
pronounced and our calculations based on small cluster diagonalization may
be less accurate.

In conclusion, we have performed cluster exact diagonalization based
LDA+DMFT calculations for quasiparticle spectra in selected transition metal
oxides. Low energy excitations were compared in details with the ARPES
experiments including the ZR states in selected HTSCs, as well as doublet,
triplet, quadruplet and quintet states in NiO, CoO, FeO, MnO, respectively.
Generally good agreement between the theory and experiments was found using
single site DMFT approximation for classical Mott--Hubbard systems, while
for the case of the cuprates, going beyond the single impurity is essential.
The reduction of the spectral weight in the ZR band when approaching the $%
\Gamma $ point was monitored for all materials and explained based on the
tight--binding argument.

The authors acknowledge useful conversations with G. Kotliar and G.
Sawatzky. The work was supported by NSF DMR Grants No. 0608283, No. 0606498.
X.G.W. acknowledges support from National Key Project for Basic Researches
of China (2006CB921802), and Natural Science Foundation of China under Grant
No. 10774067.

\end{document}